***Electrostatic edge effects versus edge doping in graphene nanoribbons: a scanning gate microscopy and modelling study***

*C. Durkan, X. Liu & E. Saunders*

**Abstract.**

*The charge neutrality point (CNP) of graphene nanoribbons (GNRs) patterned by oxygen plasma etching moves towards zero as the ribbon narrows and, on 300 nm $SiO_2$, changes sign below a crossover width of ~ 50 nm: an observation previously attributed to n-type edges surrounding a p-type bulk. We test this edge-doping hypothesis in three ways. First, finite element modelling of complete GNR transistors with uniform doping shows that fringing of the gate field at the edges reproduces the trend of the measured CNP with width down to ~ 50nm without any edge doping, while producing a bias-dependent inversion of the carrier type at the edges that mimics n-type edges in scanned-probe measurements. Second, fitting the CNP of ribbons on 90 nm and 300 nm $SiO_2$ with the gate capacitance of a finite strip, which fixes the electrostatic contribution with no adjustable parameter other than the bulk doping, shows that fringing accounts for most of the width dependence and isolates the remainder as an edge charge. This is acceptor-like (0.24 e/nm per edge) in the present ribbons, which therefore have no n-type edges, and donor-like but small in devices on 300 nm $SiO_2$; no single edge-charge, field-coupled or contact-doping mechanism describes both. Third, scanning gate microscopy finds no edge contrast of the sign expected for n-type edges. The width dependence of the CNP is therefore predominantly electrostatic, and n-type edges are not an intrinsic consequence of oxygen-plasma etching. A design map shows that fringing dominates the width dependence whenever the gate dielectric is thicker than 30-50 nm, whereas in scaled high-K transistors, edge charge dominates and would shift the threshold voltage of a 10 nm ribbon by ~ 0.1-0.15 V. Fitting the CNP versus width with the strip capacitance provides a simple test for edge doping in any gated two-dimensional nanostructure.*

# 1 Introduction

There have been significant efforts over the past two decades in developing novel electronic and optical devices using graphene. However, there are a number of well-known limitations which need to be overcome with intrinsic graphene including (i) the absence of a bandgap, resulting in low on/off current ratios, rendering it impractical as a candidate material for transistors and (ii) the absence of current saturation and the low density of states near the Dirac point, which limit gain and on-state performance. However, the first of these issues can be addressed as a bandgap can be artificially induced via quantum confinement by constricting graphene laterally, i.e. by forming graphene Nanoribbons (GNRs) [1–3], by imposing a periodic potential [4–6], or by applying a perpendicular electric field to bilayer graphene [7].

A previous experimental study [3] on GNRs demonstrated a width-dependent shift of the charge neutrality point (CNP) from positive to negative as the width reduced below a certain value dependent on the geometry of the device, but typically in the range of a few tens of nm. This led to the hypothesis that graphene nanoribbons fabricated using oxygen plasma treatment have n-type edges and a p-type bulk. In this article, we directly investigate this via a combination of Scanning Gate Microscopy and modelling. The two candidate explanations make distinct predictions. Chemically n-type edges constitute fixed doping, and should therefore produce a local n-type response at the edges that is independent of the applied biases. In contrast, purely electrostatic edge charge accumulation, as predicted analytically for gated graphene strips [8, 9] and found in calculations of nanoribbon capacitance and conductivity [10, 11], is a field effect: any edge contrast it produces should depend on, and reverse with, the gate polarity. A third discriminator, as we show below, is that fixed edge doping sets the crossover width independently of the gate-oxide thickness, whereas an electrostatic edge effect scales with it; because the electrostatic contribution can be calculated exactly for a strip above a gate, the width dependence of the CNP can be used to measure the edge charge directly, provided the bulk doping of each set of devices is also determined.

The width, edge structure and doping of a graphene nanoribbon affect the bandgap and the overall electrical properties and this can be exploited in the fabrication of devices in order to tailor device characteristics and performance. However, in order to achieve a bandgap comparable to that of Silicon, widths somewhat below 10 nm are required, mostly necessitating the use of resist-based patterning techniques. It was shown [1] that even 10-nm wide GNRs have a relatively small bandgap in the range 100-200 meV, as compared to Silicon's value of 1.12 eV. Another issue is that the fabrication methods used to create nanoribbons can introduce significant edge defects which affect the transport properties and reduce the high charge carrier mobility that is possible in graphene. It is therefore highly unlikely that any such device architectures will be practical to implement at scale unless the edge structure can be characterised and controlled. Nonetheless, there is much that can be learned about electronic systems at reduced lengthscales by exploring such devices, which may be applicable more broadly.

Although graphene's intrinsic transport properties are exceptional — with room-temperature mobilities up to 180,000 $cm^2V^{-1}s^{-1}$ on hBN [12], mean free paths in excess of 1 μm [13], and breakdown current densities of order $10^8$ $A/cm^2$ in GNRs [14], comparable to those of metallic nanowires [15] — in practice they are highly sensitive to disorder: rippling, substrate impurities and charge inhomogeneity give rise to electron and hole puddles which, among other things, account for the finite minimum conductivity observed at the Dirac point in real devices [16].

The surface of graphene is typically covered with adsorbates comprising contaminants from the atmosphere and chemical residues from device fabrication. These defects have the effect that graphene is often unintentionally doped (mostly p-type), with typical measured carrier densities in the range $10^{11}$ – $10^{12}$ $cm^{-2}$ [17]. The consequence of this is not only a higher conductance: careful choice of metal electrode material is required to avoid a Schottky-type contact and significant band bending. Even in the absence of any ambient contaminants, the underlying substrate can influence the doping of graphene – for instance graphene on pristine $SiO_2$ has been shown to be n-type [18].

In terms of previous work on GNRs, it has been demonstrated [1, 19] that the bandgap, $\Delta E$, scales with the width, $w$, of a nanoribbon as $\Delta E \propto 1/w$. This dependence further illustrates the expected Dirac-like Fermion nature of the charge carriers as opposed to Schrödinger particle-in-a-box behaviour which would lead to $\Delta E \propto 1/w^2$. The edge structure, i.e. armchair or zigzag is critical in determining the GNR's properties. Tight-binding calculations predict that zigzag ribbons have no bandgap whereas armchair ribbons can [20], while first-principles calculations show that both acquire gaps that scale inversely with width once edge magnetism and edge passivation are included [19, 21]. Due to their computational cost, however, first-principles calculations of this kind are largely limited to ribbons a few nanometres wide, narrower than the lithographically patterned ribbons studied here. For this reason, we present a mesoscopic continuum model which is able to reproduce many of the key experimental features and which is useful for interpreting our results. This model is implemented in COMSOL Multiphysics [22], where drift-diffusion equations can be used to model the transport in graphene devices as long as the channel length is longer than the mean free path [23], i.e. where the transport is not ballistic.

Previous experimental work [3] investigated the effect of the width of GNRs on their resistivity. The results showed the trend of the resistivity increasing as the width decreases, particularly below widths of around 50 nm. The rate at which resistivity increases with decreasing width was shown to be much faster than for the case of metallic nanowires where resistivity scales as $1/w$ once the width decreases below the bulk mean free path [24]. It was shown that this difference in behaviour is due to a combination of (i) fabrication methods causing a high degree of edge disorder and therefore primarily diffuse scattering at the edges and (ii) deviations in the bulk band structure starting to become noticeable, with the emergence of a bandgap and subsequent suppression of carrier density. As width reduces, edge scattering effects become more prominent and so the mean free path reduces and the resistivity increases. It was also observed that the CNP switched polarity at a width of around 50 nm where the ribbons are predominantly p-type above 50 nm and n-type below 50 nm. The precise value of the width, which we will call the "crossover width" at which this switch occurs may depend on the underlying oxide thickness. From reference [3] (where devices were prepared using exactly the same process conditions and in the same lab as for the study presented here), this switch occurs at 50nm width for a 300 nm thick oxide (35–40 nm for the as-fabricated devices measured in vacuum; Figure 10), whereas in this study, on a 90 nm thick oxide, no switch is observed down to the narrowest ribbon measured (27 nm). One would expect that this crossover is associated with changes in charge carrier distribution, which is something that can be visualised by using Scanning Gate Microscopy (SGM) measurements using an Atomic Force Microscope (AFM). Jalilian *et al* used SGM to spatially map puddles of electron and hole concentrations in a Graphene field effect transistor sample, with a spatial resolution of order 1 μm [25]. These charge puddles were likely from dopant sources such as metal contacts, surface contaminants and edge effects.

In this article, we report on the use of much higher resolution SGM measurements to map variations in the charge carrier density in graphene nanoribbons with the aim of visualising whether there is a p-type bulk and n-type edges that would explain the observed CNP characteristics. The lateral resolution here is set by the tip radius and tip–sample separation and is of order a few tens of nanometres, more than an order of magnitude finer than in reference [25].

The contributions of this work are fourfold. (i) Finite-element modelling of complete GNR transistors, including the scanned tip, shows that fringing of the gate field alone reproduces the trend of the CNP with width, and that it produces an apparent, bias-dependent n-type edge which could be mistaken for chemical doping in scanned-probe measurements. (ii) A scaling argument identifies the gate-oxide thickness as a discriminator between electrostatic and chemical edge effects that can be applied to existing transport data: fixed edge doping sets the crossover width independently of the oxide, whereas the electrostatic contribution depends only on the ratio of ribbon width to oxide thickness; applied quantitatively, with the exact gate capacitance of a finite strip, it separates the electrostatic contribution to the CNP from the edge charge, which we find to be acceptor-like in the present ribbons and donor-like, but small, in the earlier devices of reference [3]. (iii) SGM with a resolution of a few tens of nanometres finds no n-type signature at the ribbon edges. (iv) A design map identifies when fringing or edge charge controls the width dependence of the threshold voltage, from research devices on thick $SiO_2$ to scaled high-κ transistors. Together these show that n-type edges are not needed to explain the width dependence of the CNP, and are absent in the present devices.

## 2 Apparatus and Experimental Techniques

In scanning gate microscopy, an electrically conductive AFM tip is used as a local top gate to map the conductance characteristics of an underlying sample. A voltage bias is applied to the tip which leads to a modification of the local carrier density in the sample immediately below it via the field effect. It is important to note that the tip and sample are not in Ohmic contact (either by operating in non-contact mode or by having a thin insulating layer on the surface) , so there is no direct transfer of charge from one to the other. The resultant changes in local carrier concentration alter the conductance of the sample, which can be measured as a change in current in a Field-effect Transistor (FET) configuration, i.e. where Source and Drain metal contacts are added across the sample. Whether the current increases or decreases is determined by whether the tip enhances or depletes the hole or electron carrier density, allowing the local carrier type to be determined. A back gate is used to set the global operating point of the sample, and hence the magnitude and sign of, the change in current induced by the scanning top gate (the AFM tip). There are two modes in which SGM can be used – (i) fixed bias and (ii) fixed-tip, both of which were simulated in this work; fixed-bias mode was employed experimentally. In these modes, changes in the current passing through the sample are measured as a function of the tip bias or position. The spatial resolution of SGM will depend on the tip size and shape, the tip voltage and the tip-sample distance, as well as on the diffusion length of the charge carriers in the sample, similar to what is seen with scanning spreading resistance microscopy [26].

In fixed bias mode, the AFM operates in tapping mode, and the tip is held at a fixed DC voltage bias. For each tip location, the current flowing through the sample is measured and logged. A current map is therefore collected at the same time as the topography and the two can be correlated. This is generally repeated for a number of different tip voltages and average tip-sample distances.

In fixed-tip mode, the AFM tip is held (usually a few tens of nm) above the sample at a specific location and the current through the sample is measured as a function of the tip voltage. As in fixed-bias mode, the tip voltage leads to an enhancement or depletion of the charge density in the sample. The difference between the tip voltage and the back-gate voltage at which the current through the sample (from Source to Drain) reaches its minimum value is the local CNP. The polarity of this local CNP tells us whether the area directly under the tip is p-type or n-type so by performing this SGM measurement at different points in the sample, information about the spatial variations in the majority carrier types can be gained.

The setup that was used, as well as the device architecture are shown in Figure 1(a). The samples were mounted in a miniature probe station that was built on the AFM sample scanning stage. The AFM tip (top gate) voltage, $V_{tg}$, and the back gate voltage, $V_{bg}$, were independently set. Using conductive BeCu probes with a nominal radius of 5 µm, the GNR was grounded at one end and a small voltage, $V_S$ was applied to the other end. The resulting current flowing through the sample, $I_s$, was collected using a Femto DLPCA-200 variable gain, low noise current-voltage converter, the output of which was fed to the input of an analog to digital converter (ADC) in the AFM controller. This was used with a transimpedance gain of $10^7$ V/A and a 50 kHz bandwidth.

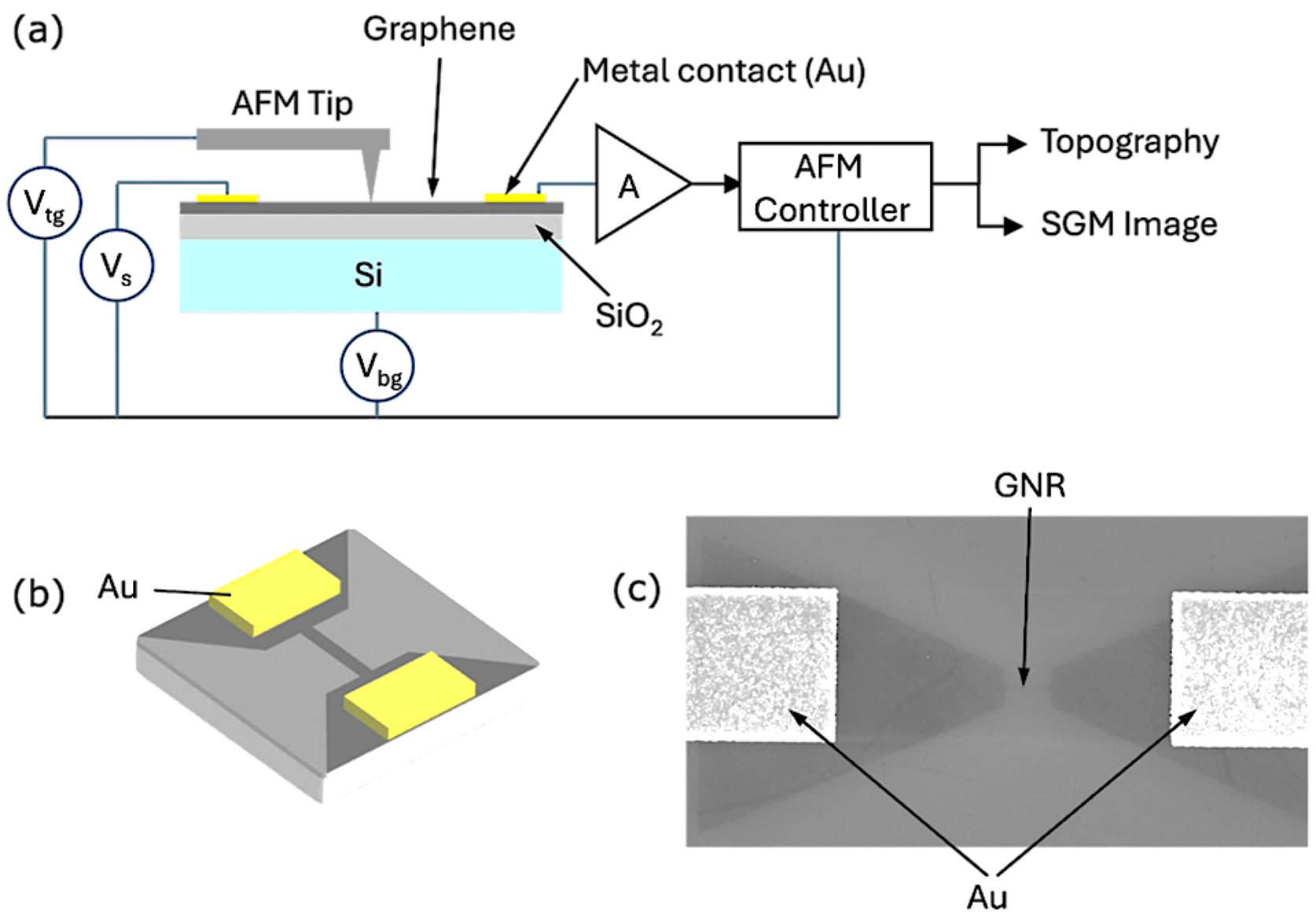

*Figure 1: (a) Experimental set-up for Scanning Gate Microscopy (SGM) where $V_{tg}$, $V_s$, and $V_{bg}$ are the top gate voltage (applied to the tip), source voltage and back gate voltage and "A" is a transimpedance amplifier; (b) device layout – the Au is deposited directly on graphene with a Cr adhesion layer, and creates the source and*

*drain contacts; (c) example device with a 15 nm wide GNR. The devices are symmetric, and the source and drain contacts are on the left and right, respectively.*

A Park Systems XE-100 AFM was used for these experiments. This is a sample-scanning system where the sample stage has sufficient access to allow for a miniature probe station to be mounted on it. The optical Microscope of the AFM was used with a 5× objective lens which combined with the other optical components gives a magnification of 500×. This was used to locate the AFM cantilever and position it above the relevant GNR as well as to position the probes on the source and drain contacts. The AFM tips used were model Multi75E-G with a nominal stiffness of 2.8 N/m, resonance frequency of 75 kHz and with an electrically conductive coating of 5 nm chromium and 25 nm platinum to give a radius of curvature at the tip apex of approx. 25 nm.

The GNRs were fabricated using the multistep process outlined in reference [3], and all samples were prepared with an alumina passivation layer approximately 7.5 nm thick. This prevents deterioration over time and ensures there are no accidental short-circuits between the tip and sample. The fabrication involves the use of an $O_2$ plasma to remove any unwanted graphene, and it is known that this will lead to a variety of C-O species at the edge, and typically leads to edges being doped p-type [27]. After fabrication and before being placed in the AFM, the transconductance of the GNRs was measured using a Keithley 4200 Source measure unit with a vacuum probe station (Lakeshore) operating at room temperature and a modest vacuum of $3x10^{-4}$ mbar. It is worth noting that for the work presented in [3], the underlying oxide had a thickness of 300nm, whereas for the devices presented here, the oxide was 90 nm thick. As discussed below, for the same bulk doping, this is expected to reduce the crossover width at which the CNP appears to switch sign if the effect is electrostatic, but not if it is due to fixed edge doping. The channel length was 600 nm for all devices, both here and in [3]. An illustration of the device layout is shown in Figure 1(b) indicating the Source and Drain contacts, and an SEM image of a device is shown in Figure 1(c) where the 5nm Cr/25 nm Au Source (left) and Drain (right) contacts can be clearly seen.

# 3 COMSOL Model

In order to be able to model the coarse electrical behaviour of the GNR devices, the semiconductor module of COMSOL was used. Given that COMSOL does not have a built-in model for graphene and an atomistic simulation would be too computationally costly, it was modelled as a thin 3D semiconductor with a near zero bandgap. We solved the Poisson equation and concentrated on the overall trends of the mesoscopic behaviour without explicitly including any quantum components. We would expect the model to be reasonably accurate down to widths in the range 10-20nm below which quantum effects and bandgap variations are known to become significant. We have also not taken account of the atomic nature of the edges, nor whether they are armchair or zig-zag. As the model contains no fixed edge charge and no local work-function offsets, it isolates the purely electrostatic contribution of the device geometry to the CNP.

The transport physics is based on the drift-diffusion (DD) model, the validity of which will start to break down for GNR lengths shorter than the mean free path due to the appearance of nonstationary and/or quasi-ballistic transport effects. However, some models suggest that by assuming a modified gate-length-dependent saturation velocity, DD provides reasonable results for smaller lengths [28]. Even without this modification, the DD approach is considered a reasonable guide even in the presence of nonstationary and quasi-ballistic transport in short-channel MOSFETs since it correctly accounts for both device geometry and electrostatics. Nonetheless, the devices we reported on in reference [3] are sufficiently large that this modelling approach should be appropriate.

A summary of the material properties used for modelling the graphene is given in Table 1. The dielectric constant of 8 was assumed based on calculations of the microscopic dielectric permittivity of graphene [29]. Similarly, the mobilities were determined experimentally from conductance measurements on the GNRs [3]. The bandgap was set to a constant value less than $k_BT$. This approach was necessary as if the value 0 is used, COMSOL was unable to converge on a solution. The acceptor doping level, $N_A$ was chosen to give good agreement with measured current-voltage characteristics for bulk graphene devices (we tested graphene with lateral size 5 μm x 25 μm). The sheet density of $1\times10^{12}$ cm$^{-2}$ corresponds to a volumetric acceptor density of $1\times10^{19}$ cm$^{-3}$ for the 1 nm-thick layer used in the model.

| **Property** | **Value** |
|---|---|
| Relative Permittivity | 8 |
| Bandgap | 5 meV |
| Electron Mobility | 3000 cm$^2$/Vs |
| Hole Mobility | 3000 cm$^2$/Vs |
| Acceptor dopant concentration, $N_A$ | $1\times10^{12}$ cm$^{-2}$ |

Table 1 - Graphene Material Properties used in COMSOL simulations

The DoS was taken, as a pragmatic approximation, from the built-in Si material with Nc and Nv reduced by a factor of ten, chosen so that the model reproduces the measured carrier density at typical gate biases. Although the true DoS of graphene is linear in energy, the trends reported here are governed by the device electrostatics rather than by the detailed form of the DoS. To avoid voltage offsets from work function differences, the contacts were set to have an equal nominal work function to graphene. The justification for this comes from the fact that experimentally, the contacts are clearly Ohmic with no evidence of any Schottky-like behaviour. Quantum capacitance has similarly been neglected; for the 90 nm oxide used here the geometric gate capacitance is small compared with the quantum capacitance of graphene at the relevant carrier densities, so the associated series correction is minor [30].

The geometry of the modelled system can be seen in Figure 2. The graphene layer is 1nm thick which was chosen as a suitable minimum value due to the constraint it puts on the size and therefore number of elements needed in the mesh. The AFM tip was modelled as a sphere with a diameter of 35 nm. It was treated as an internal boundary with an applied electric potential rather than as a metal or semiconductor domain in order to reduce the model complexity. In reality the tip cone and cantilever contribute an additional weak, long-range background gate; the strong spatial localisation of the measured response (Figure 11) indicates that the sphere approximation captures the dominant, local part of the tip field.

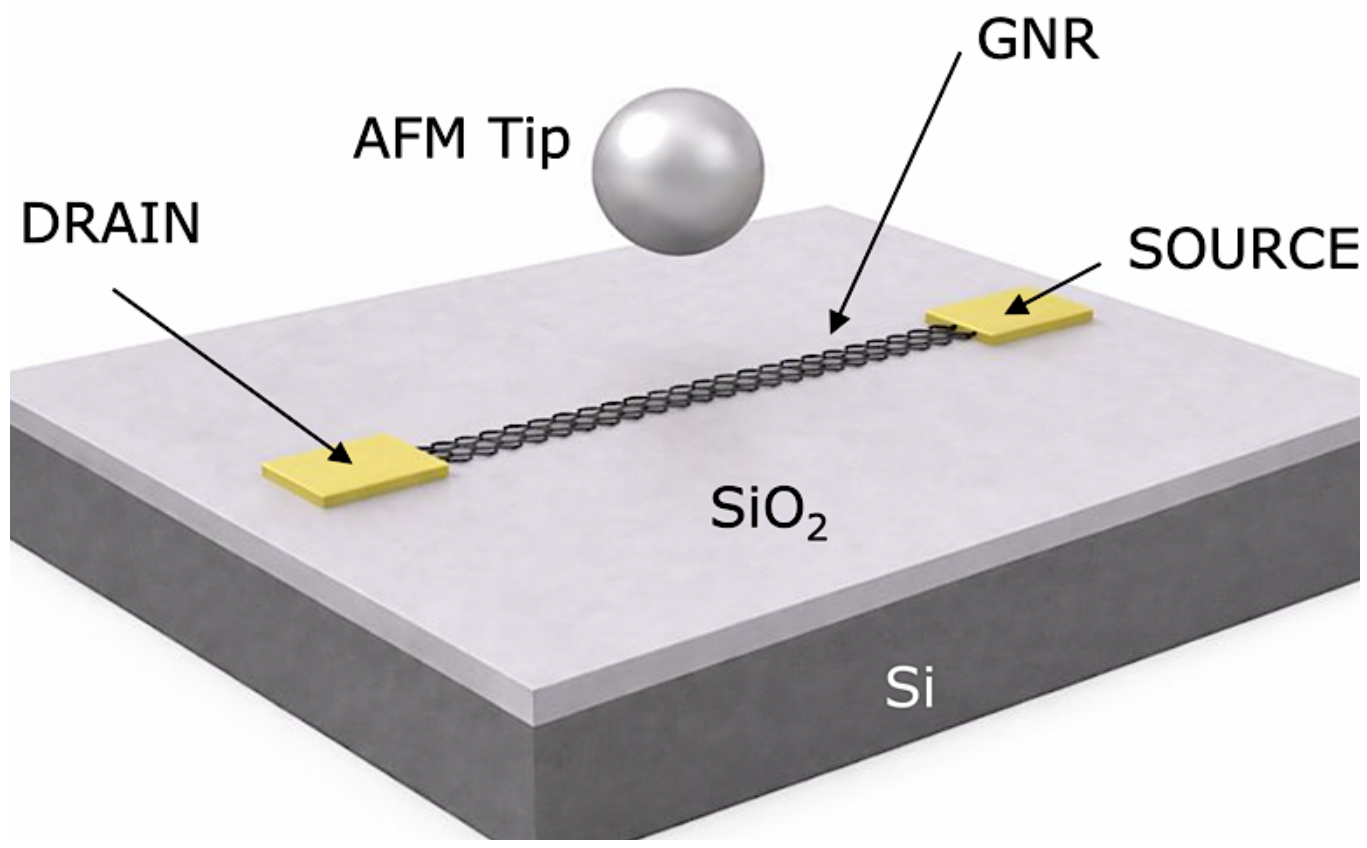

Figure 2: Schematic of the system modelled in COMSOL, where the AFM tip is approximated by a sphere for simplicity.

An infinite-element domain was used at the majority of the boundaries to minimise the variation in results with modelled domain size. Instead of modelling doped Si, the substrate/back-gate was modelled as a boundary at constant electric potential. This approach was taken as the semiconductor regions are the most computationally intensive part of the model for achieving solution convergence.

The drain and source terminals to the GNR were modelled as metal contacts to reduce the physical size of the model, instead of the graphene wedge structure present in the real devices (see Figure 1(b)). Despite the fact that the nominal work functions are equal, this still leads to carrier injection and some band-bending in the GNR within the vicinity of the contacts which then imposes a minimum length limit on the GNRs to be simulated to avoid behaviour being significantly impacted by this region. The region with band-bending was seen to extend out to around 10 nm from the contact, so the minimum length GNR which was simulated was chosen to be 100 nm. Increasing the length of the GNR increases the number of elements required to produce a suitable mesh. Given that the computational cost of the simulations scales with the number of degrees of freedom in the problem, which is proportional to the number of elements used, the maximum length (or width) of ribbons we investigated was 400 nm.

It is also worth noting that the element size in the graphene should be smaller than the Debye (Thomas-Fermi) screening length as the graphene is being modelled as a semiconductor. This is automatically managed within COMSOL and is of order 5 nm for the chosen parameters in Table 1.

# 4 Results and discussion

**COMSOL Simulations**

The first step was to simulate the field-effect conductance of a device in the absence of an AFM tip, as shown in Figure 3. The back-gate voltage was swept for a number of GNR widths from 10-400 nm. The oxide thickness used was 90 nm and the same doping concentration of $10^{12}$ cm$^{-2}$ was used for each device, as per the figures in Table 1. Similar results are seen with a reversed back-gate voltage polarity if n-type doping of the same concentration is modelled. For computational efficiency, this set of simulations used different lengths for different widths, so to facilitate comparison, conductivity has been plotted instead of current.

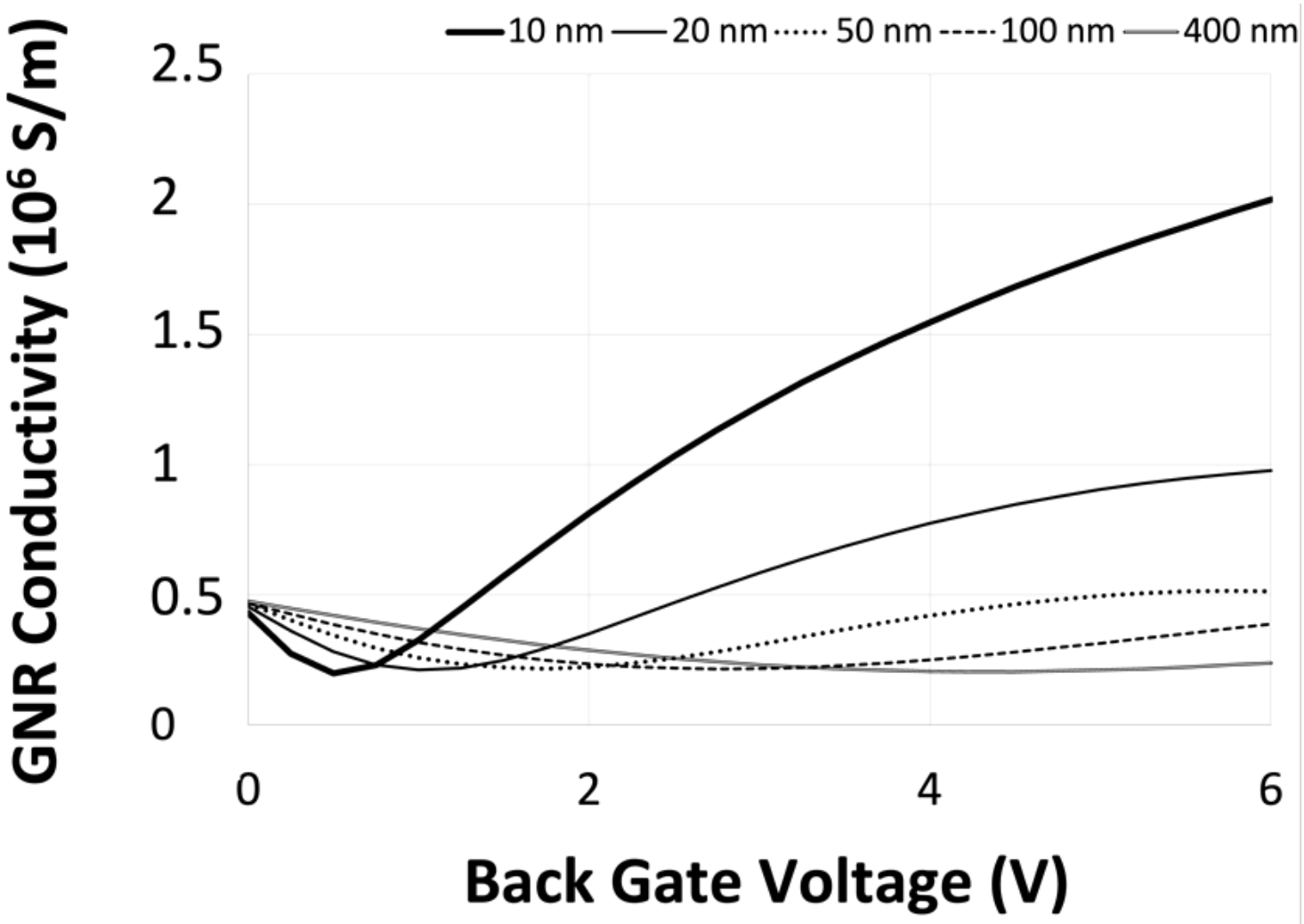

*Figure 3: Simulated global gate effect with varying GNR width, showing that the Charge Neutrality Point (CNP) shifts towards zero as GNR width is reduced.*

As is clear from this, the CNP shifts towards lower voltages as the width is reduced. In broad terms, this is similar to what was experimentally observed [3], although in the simulations, there is no evidence of a crossover, i.e. the sign of the CNP does not change. Given that the edges are modelled as having the same doping density as the bulk of the GNR, the observed shift of the CNP in Figure 3 is therefore not consistent with the hypothesis from reference [3] that the GNR edges are n-type. Instead, what we are observing is the effect of fringing of the electric field. Gate-induced charge accumulation at the boundaries of a graphene strip was predicted analytically by Silvestrov and Efetov [8] and by Fernández-Rossier *et al* [9], and appears in calculations of nanoribbon capacitance [10] and strip conductivity [11]; the present model extends this picture to the full device geometry, doping, contacts and scanned top gate of our experiments. These results can be explained by thinking of the GNR and back gate as the plates of a parallel plate capacitor, where fringing becomes significant when the lateral size of one or both of the plates becomes comparable to the distance between them. For electrodes which both have a large lateral extent compared to the separation, the electric field between them is essentially uniform. However, when the separation becomes significant compared to the lateral size of the electrodes, the fringing of the electric field at the edges of the plates can dominate the behaviour.

For the GNR-FETs, this has the consequence that the CNP of wider ribbons such as the 400 nm wide GNR in Figure 3 converges to that seen for bulk devices.

In any FET, it is the gate electric field rather than the gate bias that determines the device behaviour. As a result of fringing, the field distribution is significantly altered in small GNRs relative to larger ones, leading to a different distribution of charge and ultimately, a different conductivity.

The simulations show that the field enhancement around the edges is large enough for the CNP to fall as the width is reduced, as observed experimentally (Figure 9), i.e. the maximum depletion of the charge carriers can be achieved for a lower voltage for narrower GNRs. The field resulting from two different GNR widths can be seen in Figure 4 where we show a cross-section taken through the GNRs halfway along their length. In this figure, we show a zoom-in of the GNRs with widths of (a) 25 nm and (b) 400 nm.

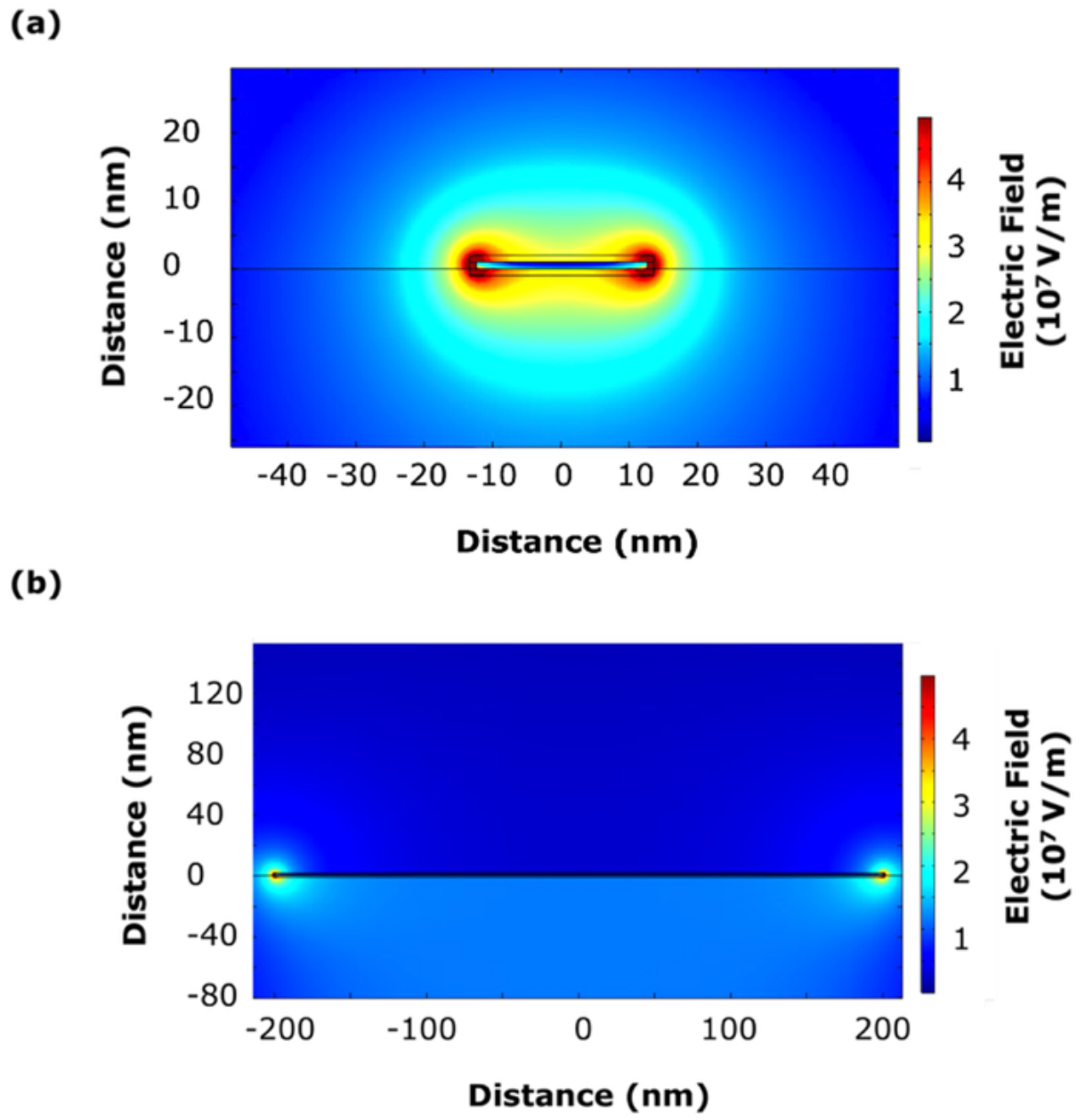

*Figure 4: Electric field strength in cross-sections of GNRs (looking along the length axis) of width (a) 25 nm and (b) 400 nm for a back-gate bias of 1 V and an underlying $SiO_2$ layer of thickness 90 nm.*

The differing degree to which fringing occurs in the two GNRs is immediately apparent, with strong field enhancement evident in the narrower one. It can also be seen that the field enhancement originates at the edges, where there are high spatial variations in potential by virtue of the geometry. The narrower GNR can be seen to have a greater average field than the wider GNR due to the more prominent fringing, therefore modulating the carrier density more strongly for the same bias, especially at the edges due to the high spatial curvature there. As can be seen in Figure 5, the effect of this field enhancement is that the edges have a higher concentration of electrons than the bulk and effectively behave as n-type. However, this fringing effect does not cause the CNP to switch sign, so it cannot fully account for the observations in reference [3]. The inversion of the charge carriers at the edges is therefore entirely a field effect and is dependent on the applied bias, so cannot be considered to be equivalent to doping.

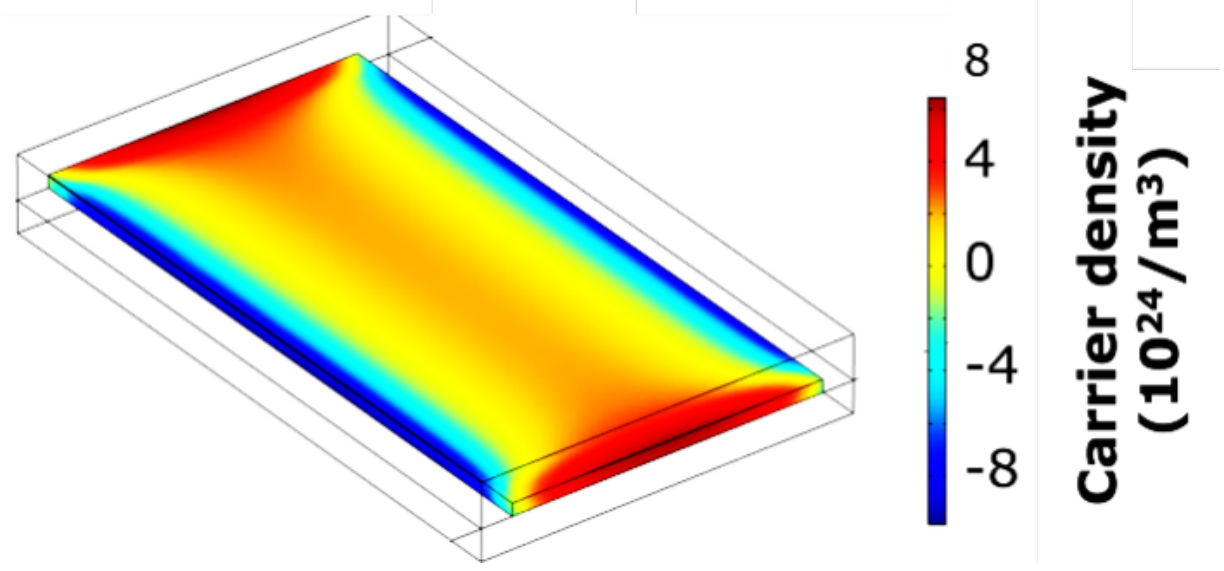

*Figure 5: Signed carrier distribution for GNR of width 25 nm at back-gate bias of 1 V. The red areas indicate majority p-type carriers and the blue areas indicate majority n-type carriers.*

We then investigated the expected result of the two SGM modes discussed above, for the case of a 50 nm wide GNR with an AFM tip present, which can act as a scanning top gate. A back-gate bias of either +1 V or -1 V was used. For the fixed tip position simulation, the tip was positioned in the centre of the GNR with a 10 nm gap between the bottom of the AFM tip and the top of the oxide surface. For the fixed bias simulation, the tip was kept at the same height but with a bias of either +1 V (the local CNP voltage when positioned over the centre of the GNR), +2 V or +4 V, and the centre of the tip was then moved laterally across the GNR starting from 130 nm from the centre (105 nm from the edge) to the opposite side, passing above the centre of the device, while calculating the drain current at each position. The predicted variation in current for the fixed tip position and fixed tip bias SGM modes are shown in Figure 6 (a) and (b), respectively.

(a) (b)

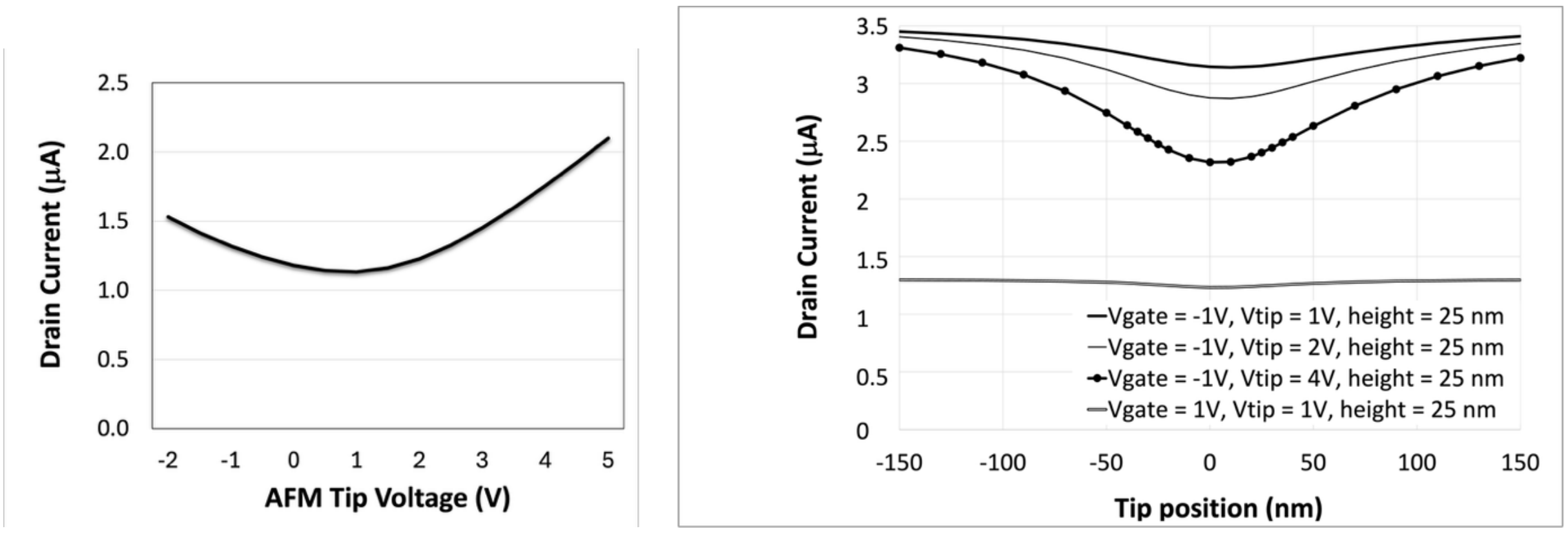

*Figure 6(a): SGM simulation of fixed tip mode for the tip above the GNR centre showing the effect of sweeping the tip bias on the GNR current, (b) SGM simulation of fixed bias mode with fixed tip and back-gate biases, where x = 0 nm is the centre of the GNR (width 50 nm)*

These plots show that the order of magnitude maximum variation in current to be expected using the SGM technique ranges from a few per cent at a tip bias of +1 V to approximately 10% at +2 V and 30% at +4 V (Figure 6(b)). However, the simulated nanoribbon here is 100 nm long due to computational constraints whereas the actual GNRs that we fabricated are around 600 nm long. Therefore, the simulations would be expected to overestimate the gate effect from the tip by a factor of around 6. Scaling the +2 V case accordingly predicts a change of order 1.5–2% for a 600 nm long ribbon, which can be compared directly with the measurements below.

The average height of the tip above the surface and the tip radius both influence the magnitude of the corresponding electric field in the GNR, and therefore have a significant effect on the current, as shown in Figure 7. The simulation shows the expected variation in current for a 15nm radius tip with a 10nm gap and then a 25nm radius tip with 10nm and 20 nm gaps. A larger tip gives rise to a larger change in current, but the spatial resolution is clearly lower, as the edge sharpness is reduced when compared against the smaller tip. Increasing the gap between the tip and sample also has the effect of reducing the overall change in current and smears out the edge even further. In other words, to be able to observe the fringing-induced carrier distribution at the edge, the AFM tip will need to be as small as possible and as close as possible to the surface of the GNR to create a highly-localised field, entirely in line with what one would expect. The value of the simulations is that it allows us to design the experiment accordingly, which we will now consider.

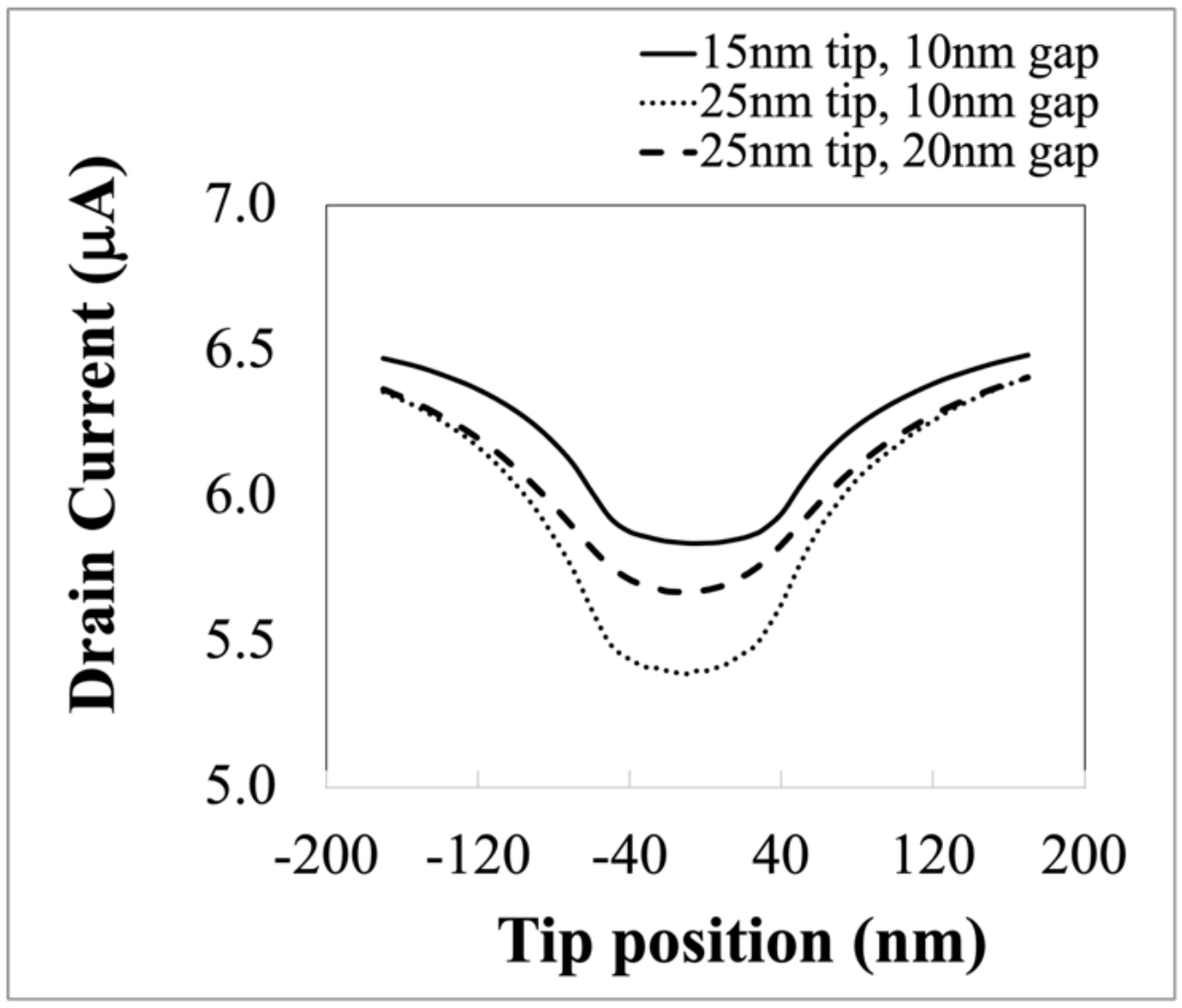

*Figure 7: COMSOL simulation of drain source current results with varying AFM tip radius and two different tip heights where x = 0 nm is the centre of the GNR.*

In the next section, the correlation between experimental results and the above simulations will be examined.

**Experimental results**

It has been seen previously that GNRs made using the process employed in this work typically appear to switch their majority dopant type at a width of around 50 nm, with GNRs seemingly being n-type below 50 nm and p-type above it. The GNRs used here were deposited on a 90 nm oxide layer as opposed to the 300 nm one used in [3] so we would expect to see the corresponding shift in CNP to only become apparent for smaller widths. As a rough estimate, based on geometric scaling, the crossover point should be of order $50 \times {}^{90}/_{300} = 15$ nm. This estimate is based on the fact that as Laplace's equation contains no intrinsic lengthscale, the electrostatic contribution to the CNP, normalised to its wide-ribbon value, only depends on the ribbon width $w$ and the oxide thickness $d$ through the ratio $w/d$ which is due to fringing, apart from corrections from the much shorter lengths in the problem (the screening length in the graphene, the 7.5 nm passivation layer and the tip–sample spacing).

The expected trend of reducing CNP with reduced GNR width is indeed apparent from Figure 8 where we show the results of a back gate voltage sweep for GNR FETs with widths ranging from 27nm to 320nm. No switch in polarity is observed down to the narrowest ribbon measured (27 nm), consistent with the estimated crossover width of ∼15 nm for this oxide thickness. The significance of this for the edge-doping hypothesis is discussed below. For comparison, the simulated values of CNP, as extracted from figure 3 are shown in Figure 9. It is clear that the overall trends of both theory and experiment are roughly the same and that agreement is reasonable for mid-range widths, but they start to deviate noticeably for widths below around 50 nm. The simulation predicts a stronger dependence on width than is observed experimentally. The experimental CNP is almost constant between 60 nm (2.55 V) and 27 nm (2.3 V), whereas the uniformly doped model continues to fall to ≈1 V, so the deviation is in the direction of excess p-type character in the narrowest ribbons. In the linear electrostatic model, a uniform acceptor density $p_b$ contributes $ep_bw/C'(w)$ to the CNP, where $C'(w)$ is the gate capacitance per unit length, and a fixed acceptor-like charge $\lambda$ per unit length at each edge adds $2e\lambda/C'(w)$, so the ratio of measured to simulated CNP can be read as $1 + 2\lambda/(p_bw)$. Interpolating the simulated values, the ratios of ≈1.1, 1.3 and 2.3 at 120, 60 and 27 nm would correspond to an acceptor-like edge charge of order $10^6$ $cm^{-1}$ (≈0.1 e $nm^{-1}$) per edge if the whole deviation were due to edge charge; part of it may instead reflect the idealised geometry of the model. A fit that treats the fringing exactly, rather than through the simulated values, gives ≈0.24 $e$ $nm^{-1}$ per edge (Figure 10, below). This is the sign expected for the C–O groups formed at oxygen-plasma-etched edges [27]. In neither case does the deviation point towards n-type edges.

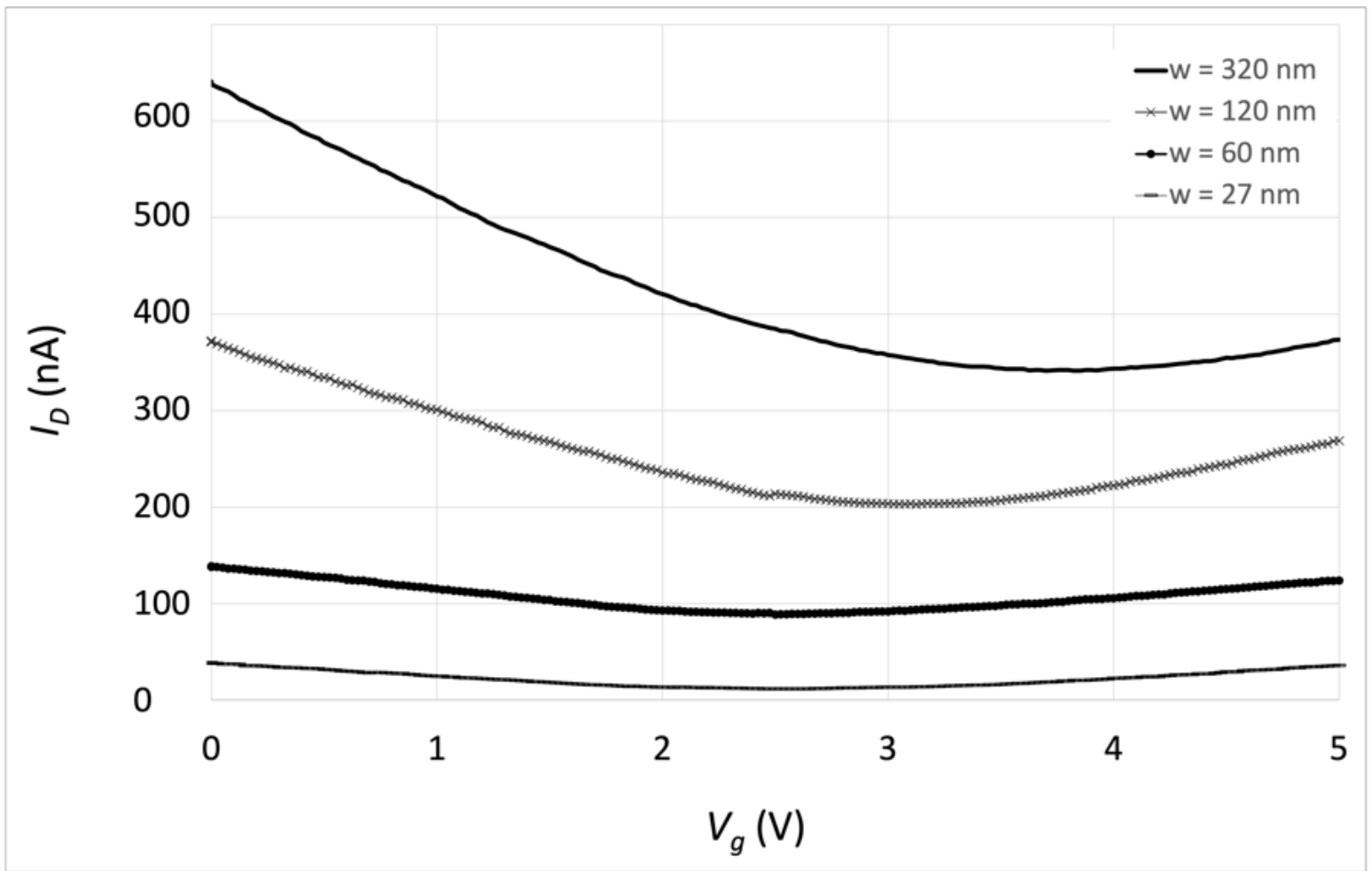

*Figure 8: Experimental Measurement of Drain Current vs back gate voltage for different GNR widths, showing the shift of CNP towards lower values as width decreases.*

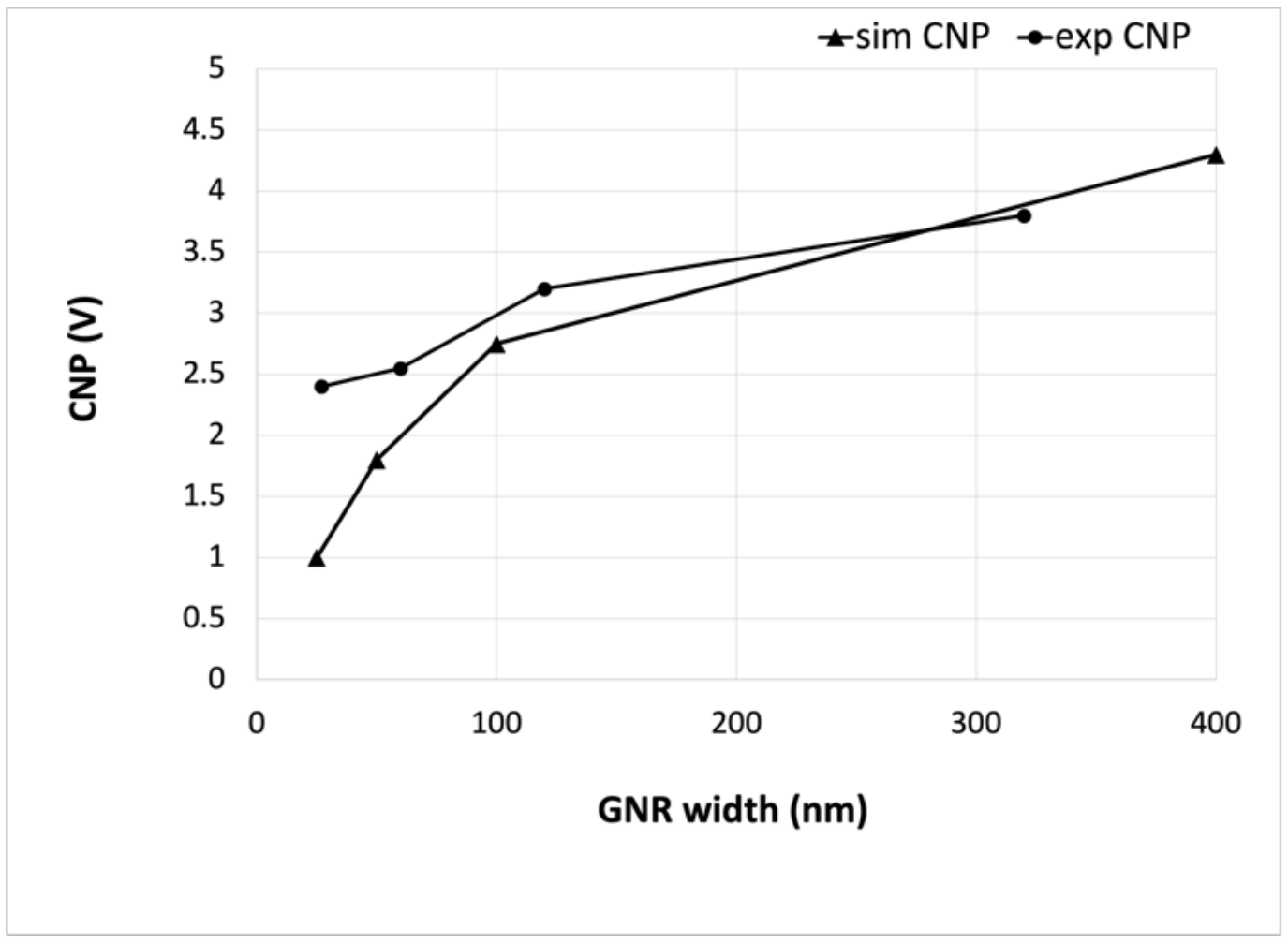

*Figure 9: Comparison of simulated (sim) and Experimental (exp) values of CNP for different GNR widths, in the absence of the AFM tip top-gate.*

The comparison between the two oxide thicknesses provides a test of the edge-doping hypothesis that does not depend on the details of the model. Suppose that the edges carry a fixed donor charge $\lambda$ per unit length and the bulk a fixed acceptor density $p_b$. For ribbons narrower than the oxide thickness the neutrality condition applies to the ribbon as a whole and, neglecting quantum capacitance and work-function offsets as in the model, the CNP is $V_{CNP}(w) = e(p_b w - 2\lambda)/C'(w)$. Fringing changes $C'(w)$, and hence the magnitude of the CNP, but not its sign, so the crossover occurs at $w_c = 2\lambda/p_b$ whatever the oxide thickness. With $w_c \approx 50$ nm on 300 nm oxide [3], edge doping of the same strength would, for the same bulk doping, make every ribbon narrower than 50 nm n-type on 90 nm oxide as well, whereas the 60 nm and 27 nm ribbons remain clearly p-type (Figure 9). The bulk doping is not the same, however. From the CNP of the widest (320 nm) ribbons, $p_b \approx 0.9 \times 10^{12}$ cm$^{-2}$ for the present devices but only $\approx 0.2 \times 10^{12}$ cm$^{-2}$ for the as-fabricated devices of reference [3] in the parallel-plate approximation (1.3 and 0.35 × $10^{12}$ cm$^{-2}$, respectively, when fringing is included). A donor edge charge of the strength needed on 300 nm oxide would therefore move the crossover on 90 nm oxide to $w_c \approx 12$ nm, below the narrowest ribbon measured, so the absence of a sign change down to 27 nm is not by itself evidence against fixed edge doping. The comparison must instead use the full width dependence on both oxides, as follows.

To make this comparison quantitative, we calculated $C'(w)$ for a zero-thickness strip on $SiO_2$ above a grounded back gate from the closed-form quasi-static expressions for a microstrip line [31]. These agree with a two-dimensional finite-difference solution of Laplace's equation to within 0.3% for all the widths and oxides used here (grid convergence ≈0.3%), and the finite-difference solution was also used to include the 7.5 nm $Al_2O_3$ passivation ($\varepsilon_r \approx 9$) of the present devices. Because the channels are only 600 nm long, comparable to twice the 300 nm oxide thickness, we also solved the three-dimensional problem of a ribbon joined at each end to tapered graphene leads as in Figure 1(c). The leads screen the gate field over a distance of order $d$ from each end, reducing the length-averaged $C'$ by 2–6% on 90 nm oxide and by 8–16% on 300 nm oxide, slightly more for narrower ribbons; this correction is included below but changes none of the conclusions. Treating the ribbon as an equipotential, $C'(w)$ fixes the electrostatic contribution $ep_b w/C'(w)$ to the CNP with no adjustable parameter other than $p_b$, so any additional width-dependent contribution can be isolated. We fitted the 90 nm data (Figure 8) together with the data of reference [3] for as-fabricated devices measured in vacuum, allowing $p_b$ to differ between the two sets of devices, with five descriptions of the additional contribution: (E) none; (A1) a fixed charge $\lambda$ per unit length at each edge, the same on both oxides; (A2) a fixed edge charge allowed to differ between the oxides; (B) an edge charge proportional to the fringing charge, as expected for traps charged by the fringing field, which adds $V_B(1 - C'_{pp}/C')$ to the CNP, where $C'_{pp} = \varepsilon_0\varepsilon_{ox}w/d$; and (C) contact-induced doping, which extends 0.2–0.3 μm into the channel on 300 nm $SiO_2$ [32] and, scaling this length with $d$, adds $V_c \cdot \min(1, 1.66d/L)$ for a channel of length $L$ = 600 nm. In our devices the metal contacts are separated from the ribbon by the graphene leads (Figure 1(c)), so (C) is an upper bound on contact effects. Here $\lambda$ is counted positive for acceptor-like (p-type) charge, so that $V_{CNP} = e(p_b w + 2\lambda)/C'(w)$. The results are summarised in Figure 10 and Table 2.

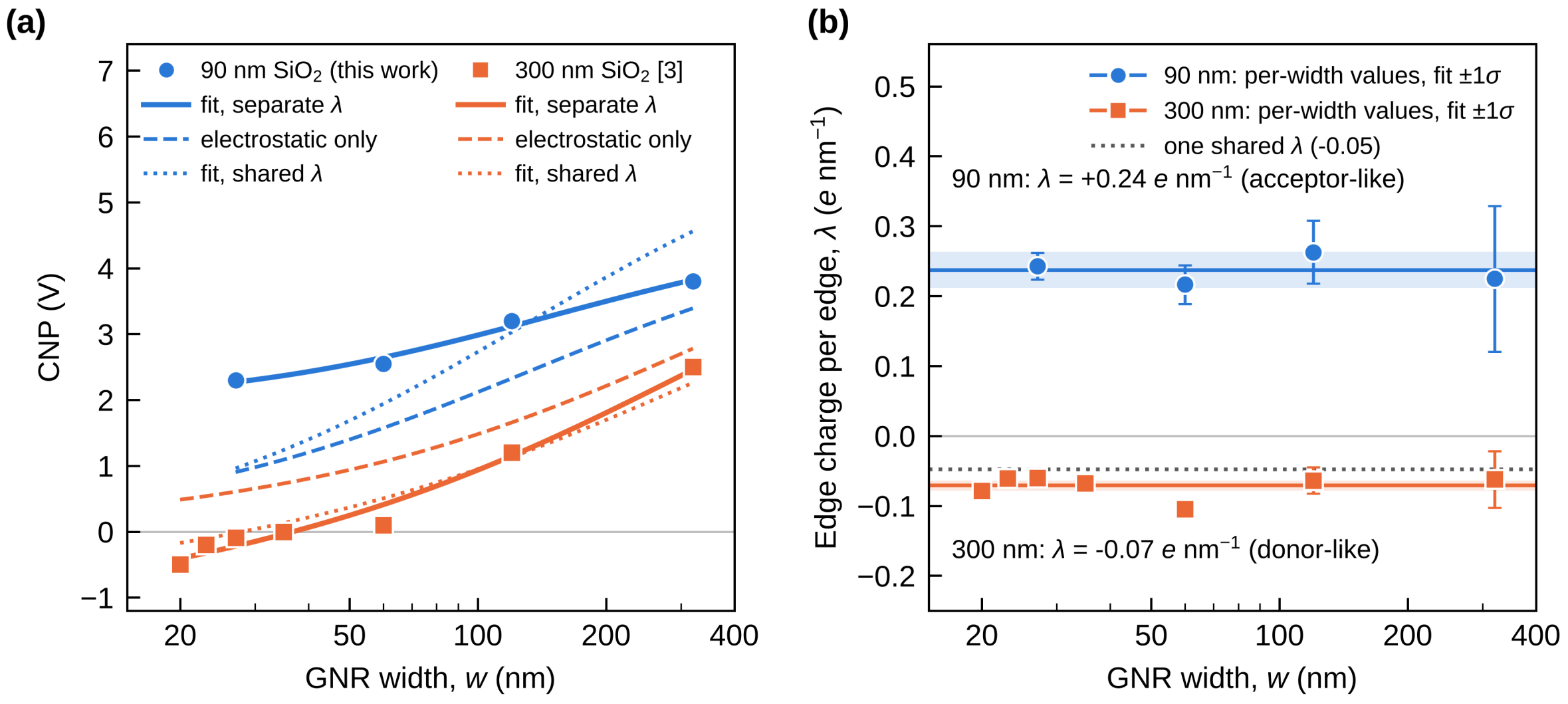

*Figure 10: (a) CNP versus ribbon width on 90 nm $SiO_2$ (this work, circles) and on 300 nm $SiO_2$ (as-fabricated devices of reference [3] measured in vacuum, squares); all channels are 600 nm long. Dashed lines: electrostatic contribution $ep_b w/C'(w)$ alone; solid lines: with a fixed edge charge fitted separately for each oxide (model A2); dotted lines: with one edge charge shared by both oxides (model A1). (b) Edge charge per edge inferred at each width, $\lambda = [C'(w)V_{CNP}/e - p_b w]/2$, using $p_b$ from model A2; error bars combine an*

*assumed ±0.1 V uncertainty in each CNP with the uncertainty in $p_b$. Solid lines and bands: fitted $\lambda \pm 1\sigma$; dotted line: shared fit.*

Electrostatics alone describes neither data set (Figure 10(a), dashed lines): relative to the electrostatic prediction, the narrowest ribbons are too p-type on 90 nm oxide and too n-type on 300 nm oxide. A fixed edge charge describes each data set well, with r.m.s. deviations of 0.07 V and 0.14 V, comparable to the uncertainty of a CNP reading, but requires $\lambda = +0.24 \pm 0.03\ e\ \mathrm{nm}^{-1}$ (acceptor-like) on 90 nm oxide and $\lambda = -0.071 \pm 0.007\ e\ \mathrm{nm}^{-1}$ (donor-like) on 300 nm oxide, with $p_\mathrm{b} = (1.16 \pm 0.06) \times 10^{12}$ and $(0.39 \pm 0.02) \times 10^{12}\ \mathrm{cm}^{-2}$, respectively. Inferred width by width, $\lambda$ shows no systematic trend on either oxide (Figure 10(b)), as expected for a fixed edge charge. A single edge charge shared by both oxides cannot reconcile them: the best compromise, $\lambda = -0.05\ e\ \mathrm{nm}^{-1}$, misses the 90 nm data by 0.83 V r.m.s., and with the donor-like charge required on 300 nm oxide the 27 nm ribbon on 90 nm oxide would have a CNP of 0.5 V instead of the measured 2.3 V. Neither the field-coupled charge (B) nor contact doping (C) does better with one shared parameter (Table 2), because each has a fixed sign whereas the additional contribution changes sign between the two sets of devices. A width-independent offset (O in Table 2) fits each oxide as well as an edge charge does (+1.7 V on 90 nm and −1.0 V on 300 nm oxide), so these data alone cannot distinguish an edge charge from, for example, a difference in gate work function or oxide charge between the two sets of wafers; in either description, however, the additional contribution has opposite signs in the two sets of devices. These conclusions are unchanged if the $Al_2O_3$ cap or the finite-length correction is omitted, or if the devices of [3] are also assumed to be capped ($\lambda$ then ranges from +0.15 to +0.26 $e$ nm$^{-1}$ on 90 nm oxide and from −0.071 to −0.080 $e$ nm$^{-1}$ on 300 nm oxide).

| **Model** | **Additional contribution** | **Fitted additional parameter(s)** | **r.m.s., 90 nm (V)** | **r.m.s., 300 nm (V)** |
|---|---|---|---|---|
| E | none | – | 0.69 | 0.57 |
| A1 | fixed edge charge, shared | $\lambda = -0.05 \pm 0.03\ e\ \mathrm{nm}^{-1}$ | 0.83 | 0.23 |
| A2 | fixed edge charge, per oxide | $\lambda = +0.24 \pm 0.03\ /\ -0.071 \pm 0.007\ e\ \mathrm{nm}^{-1}$ | 0.07 | 0.14 |
| B | field-coupled edge charge, shared | $V_\mathrm{B} = -0.4 \pm 0.4$ V | 0.84 | 0.37 |
| C | contact doping $\propto d/L$, shared | $V_\mathrm{c} = -1.0 \pm 0.5$ V | 0.79 | 0.16 |
| O | width-independent offset, per oxide | +1.7 ± 0.2 V / −1.0 ± 0.1 V | 0.07 | 0.13 |

*Table 2 - Fits of the CNP versus width on 90 nm $SiO_2$ (four widths, this work) and 300 nm $SiO_2$ (seven widths, as-fabricated devices of [3]); $p_b$ is fitted separately for each oxide in every model. The edge charge $\lambda$ per edge is positive for acceptor-like charge; where two values are given they refer to 90 nm / 300 nm oxide.*

Three conclusions follow. First, the width dependence of the CNP is predominantly electrostatic on both oxides: for the 27 nm ribbons, fringing alone lowers the CNP from its wide-ribbon value of 4.9 V to 0.9 V on 90 nm oxide, and from 5.4 V to 0.6 V on 300 nm oxide, while the edge term contributes +1.4 V and −0.8 V, respectively. Second, the edges of the present ribbons are not n-type. In either description the deviation from electrostatics is in the p-type direction, as expected for the C–O groups formed at oxygen-plasma-etched edges [27], and this conclusion depends neither on the bulk doping of the earlier devices nor on the crossover-width argument above. Third, the crossover on 300 nm oxide requires, in those devices, a donor-like contribution of only ≈0.07 $e$ $nm^{-1}$ per edge (one electron per ≈14 nm of edge), between a quarter and a third of the value estimated below from the parallel-plate argument, because fringing already brings the CNP of narrow ribbons close to zero; with $p_b$ = $0.39 \times 10^{12}$ $cm^{-2}$ it places the crossover at $|2\lambda/p_b| \approx 37$ nm, as observed. Whether that contribution depends on the oxide thickness cannot be decided from the present data, because the two sets of devices also differ in bulk doping, passivation and fabrication batch. Sample-to-sample variation is large even on one oxide: the older 300 nm sample of [3] gives $p_b \approx 2.0 \times 10^{12}$ $cm^{-2}$ and $\lambda \approx -0.41$ $e$ $nm^{-1}$, about five times larger than for the as-fabricated devices, although the ratio $|2\lambda/p_b|$ that sets the crossover width is similar (41 nm). A definitive test therefore requires ribbons from one fabrication run on two (or more) oxide thicknesses, analysed as above.

In order to explore the nature of the edges in more detail, we then carried out Scanning Gate Microscopy.

**AFM – SGM measurements**

SGM was carried out with -1V applied to the back gate. The experiments presented here were carried out on two GNRs which were nominally 120 nm wide and 600 nm long. Although these ribbons are far wider than any expected crossover width, chemical edge doping is a property of the ribbon perimeter and should therefore be present at any width (its detectability by SGM is assessed below); wide ribbons thus provide a robust testbed for the edge-doping hypothesis while avoiding the fragility of few-nanometre-scale devices. In Figure 11, we show a map of the change in current overlaid on the topography image, clearly demonstrating that there is a gate effect which is localised to the GNR.

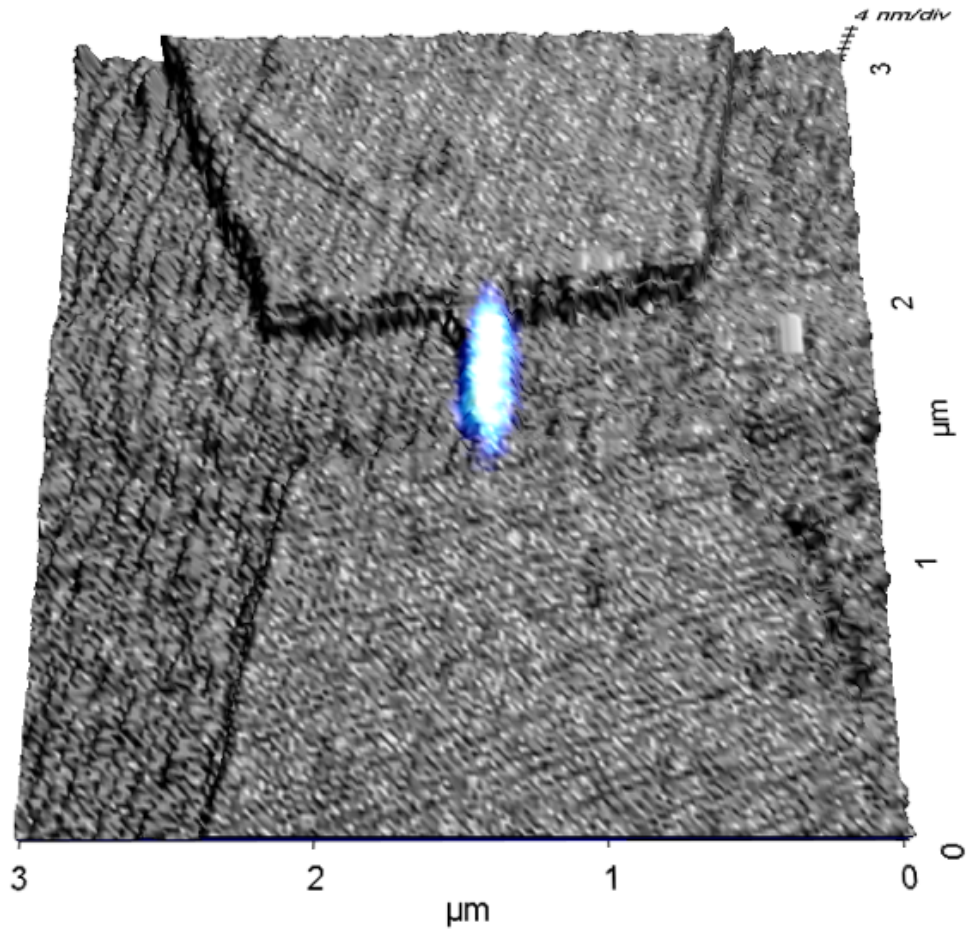

*Figure 11: 3D render of change in current on the AFM topography illustrating the spatially-localised gate effect of the tip*

In Figure 12, we plot the change in current for both GNRs, where each was measured using a different AFM tip with different apparent radius. In both figures, the baseline Drain-Source current is 600 nA. The free oscillation amplitude of the AFM tip is 33 nm and the oscillation setpoint is set at 16 nm. The average distance between the tip apex and the sample is of order 15-16 nm. In Figure 12(a), we show results from an experiment where the tip voltage was held at +2 V. There is some contamination on the GNR which is a residue from the various lithography processes used to fabricate it, hence the measured height of approximately 15 nm as opposed to the expected value of order 7.5nm. This, combined with the average tip-sample separation means we would expect the simulations for tip-sample distances in the range 15-30 nm to be appropriate. It is worth pointing out that contact-mode AFM is often employed to mechanically clean graphene surfaces and devices, but for the devices reported here this was not possible without causing irreversible damage to the GNRs given their small size.

There is no reversal of the change in current in the vicinity of the edge and therefore no evidence of a transition from *p* to *n* type. However, there is a clear step in the current change when the edge of the tip passes over the edge of the GNR in Figure 12(a), as indicated by the arrows. The overall change in current measured when the tip passes from the centre of the GNR to 100 nm from the edge is of order 12 nA, which corresponds to a change of around 2%. The edge sharpness of the topography image is approximately 25 nm, indicating a tip radius of the order 25 nm. Then in Figure 12(b), we show data for a larger tip and for two different tip voltages of 1 and 3 V. For the case of 1 V on the tip, the signal to noise ratio of the change in current is such that only a small difference can be seen as the tip passes over the centre of the GNR. In contrast, the change for a tip voltage of 3 V is significantly larger, and the total change is approximately 40 nA, or 6.7% of the baseline current. The edge sharpness of the topography image in this case is larger at approximately 40 nm, indicating a blunter tip with radius of the order 40 nm. In line with the simulation, this leads to a more noticeable but more smeared out gate effect from this tip. These magnitudes are consistent with the simulations of Figures 6 and 7 once the length scaling is applied, given that the simulated tip–sample gap (10 nm) is smaller than in the experiment: the predicted change of order 1.5–2% at a tip bias of +2 V for a 600 nm ribbon compares with the measured 2%, while the larger change of 6.7% for the blunter tip at +3 V is consistent with the simulated increase of the signal with tip radius.

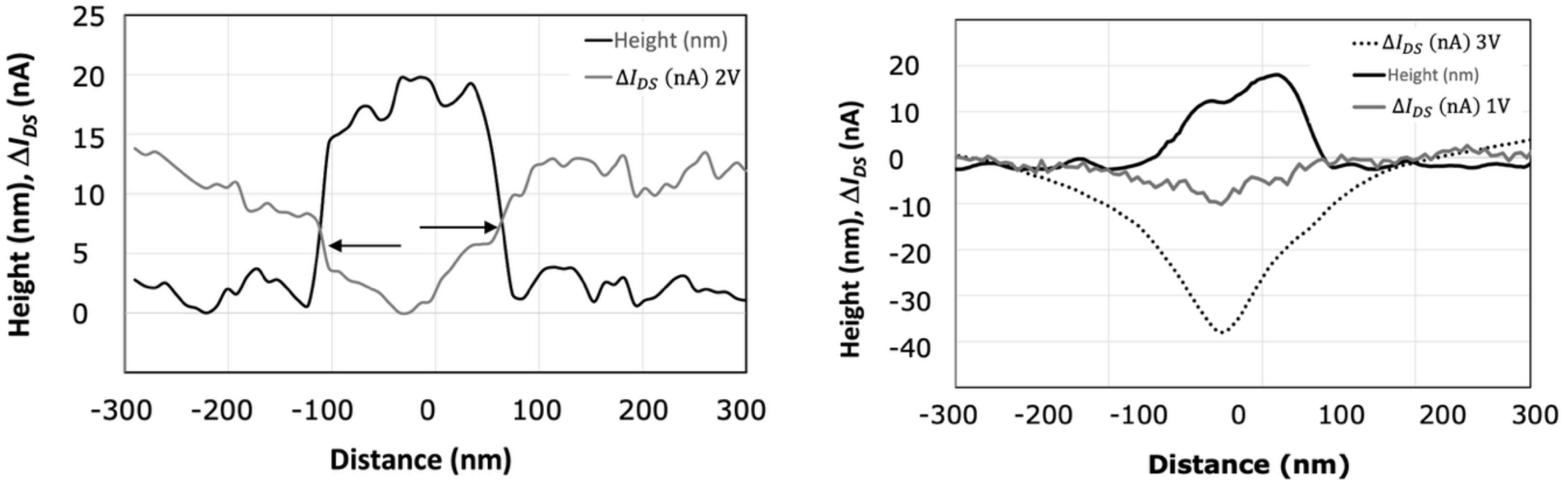

*Figure 12: Measured change in current as a function of tip position as it scans across a GNR, with topography cross-section. Both images are for different GNRs of nominal dimensions 120 nm wide and 600 nm long. (a) and (b) were taken with a tip with apparent radius of order 25 nm and 40 nm, respectively. The arrows in Fig. 12(a) indicate where there is a large change in current at the edge of the GNR.*

An objective of this study was to use SGM to probe the p-type bulk and n-type edges hypothesised by Xiao *et al* [3] as the origin of the crossover, and the expected signatures are readily stated. With the back gate at −1 V the ribbon is in its hole-accumulation regime (Figure 8), so a positive tip bias locally depletes holes and reduces the current, as is observed over the bulk of the ribbon. Over a chemically n-type edge, the same positive tip bias would instead enhance the local majority (electron) density, so the current reduction should weaken, vanish or reverse in a band along each edge — a contrast that, unlike the field-induced edge inversion of Figure 5, would not depend on the sign of the gate bias. No such feature is observed at any of the tip biases used (+1, +2 and +3 V). The one edge feature that is seen — a downward kink in the current for the smaller tip as it crosses onto the ribbon from the oxide (arrows in Figure 12(a)) — has the opposite sign to that expected for n-type edges, and is spatially localised to a region approximately 10 nm wide centred on the edge; it is most naturally attributed to the local enhancement of the electric field when the tip is in close proximity to the edge, as in Figure 4.

It is also useful to quantify the level of edge doping that the crossover hypothesis requires. For the net doping of a ribbon to invert at a width of ~50 nm with a bulk acceptor density of $1\times10^{12}$ cm$^{-2}$, the n-type charge at the two edges must fully compensate the bulk, i.e. a line density of approximately $2.5\times10^{6}$ cm$^{-1}$ per edge — equivalent to a sheet density in excess of $1\times10^{13}$ cm$^{-2}$ if confined to a 2 nm wide strip, more than an order of magnitude above the bulk doping. When fringing is treated exactly (Table 2), the requirement for the devices of reference [3] falls to $\approx 7\times10^{5}$ cm$^{-1}$ (0.07 $e$ nm$^{-1}$) per edge, or $\approx 3.5\times10^{12}$ cm$^{-2}$ in a 2 nm strip, still about nine times their bulk doping. Whether SGM can detect such an edge depends on its width rather than on its carrier density, as the tip changes the local conductivity by $e\mu\Delta n$ whatever the background density. An n-type edge region of width $w_e$ therefore contributes a response of the opposite sign whose size, relative to that of the adjacent p-type bulk, is roughly $w_e/\ell$, where $\ell \approx 50$ nm is the footprint of the sharper tip (Figure 12(a)). With that tip centred on the edge, the bulk response is ≈6 nA (half the 12 nA change between the ribbon centre and the oxide), so an n-type edge region 10–20 nm wide would produce a feature of ≈1–2 nA, of the opposite sign to the observed kink, comparable to the fluctuations of ≈1–2 nA in the current away from the ribbon. The SGM measurements therefore exclude n-type edge regions wider than ≈10–20 nm, but not a narrower strip of high density, which is why the quantitative analysis of the CNP above provides the more stringent test. Taking this together with (1) the good agreement between simulated and experimental line profiles obtained without any difference in edge doping, (2) the agreement of the CNP-versus-width trends down to widths of around 50 nm (Figure 9), and (3) the acceptor-like sign of the edge charge extracted from the CNP of the present ribbons (Figure 10), we conclude that the shift of the CNP with width is predominantly electrostatic and that the present ribbons do not have n-type edges. The small donor-like contribution that carries the CNP through zero in the devices of reference [3] remains to be identified. It is absent from the present devices, and it does not behave as charge driven by the fringing field (model B) or as contact doping (model C). Candidates that may differ between the two sets of devices include the passivation, adsorbates on the etched edges, and charge in the oxide or at its surface next to the ribbon; a size-dependent change in band structure is unlikely at these widths, since the confinement gap expected at 50 nm from the scaling of reference [1] is only 5–10 meV, well below $k_B T$ at room temperature.

# 5 Implications for scaled devices

The analysis above also indicates when edge charge matters in practice. Relative to a wide ribbon, the width-induced shift of the CNP, and hence of the threshold voltage, is the sum of a fringing term, $\Delta V_f = (e p_b t/\varepsilon_0 \varepsilon_r)(1 - C'_{pp}/C')$, where $t$ and $\varepsilon_r$ are the thickness and relative permittivity of the gate dielectric, and an edge term, $\Delta V_e = 2e\lambda/C'$. Their ratio, $R = (w/|w_c|)(C'/C'_{pp} - 1)$ with $w_c = 2\lambda/p_b$, depends on the geometry only through the fringing excess $C'/C'_{pp} - 1$, which is mapped in Figure 13(a). For $w \ll t$, $C'$ varies only logarithmically with $w$, so $R$ is almost independent of width and the boundary $R = 1$ lies at a dielectric thickness comparable to $|w_c|$: for $|w_c| \approx 40$ nm, the value found on both oxides here, it lies at an EOT of 27–40 nm for $SiO_2$ and 8–10 nm for $HfO_2$ (physical thickness 41–49 nm). Both the present devices and those of reference [3] lie inside the fringing-dominated region, which is why the electrostatic term had to be removed before the edge charge could be identified, and why fringing can masquerade as edge doping in back-gated research devices.

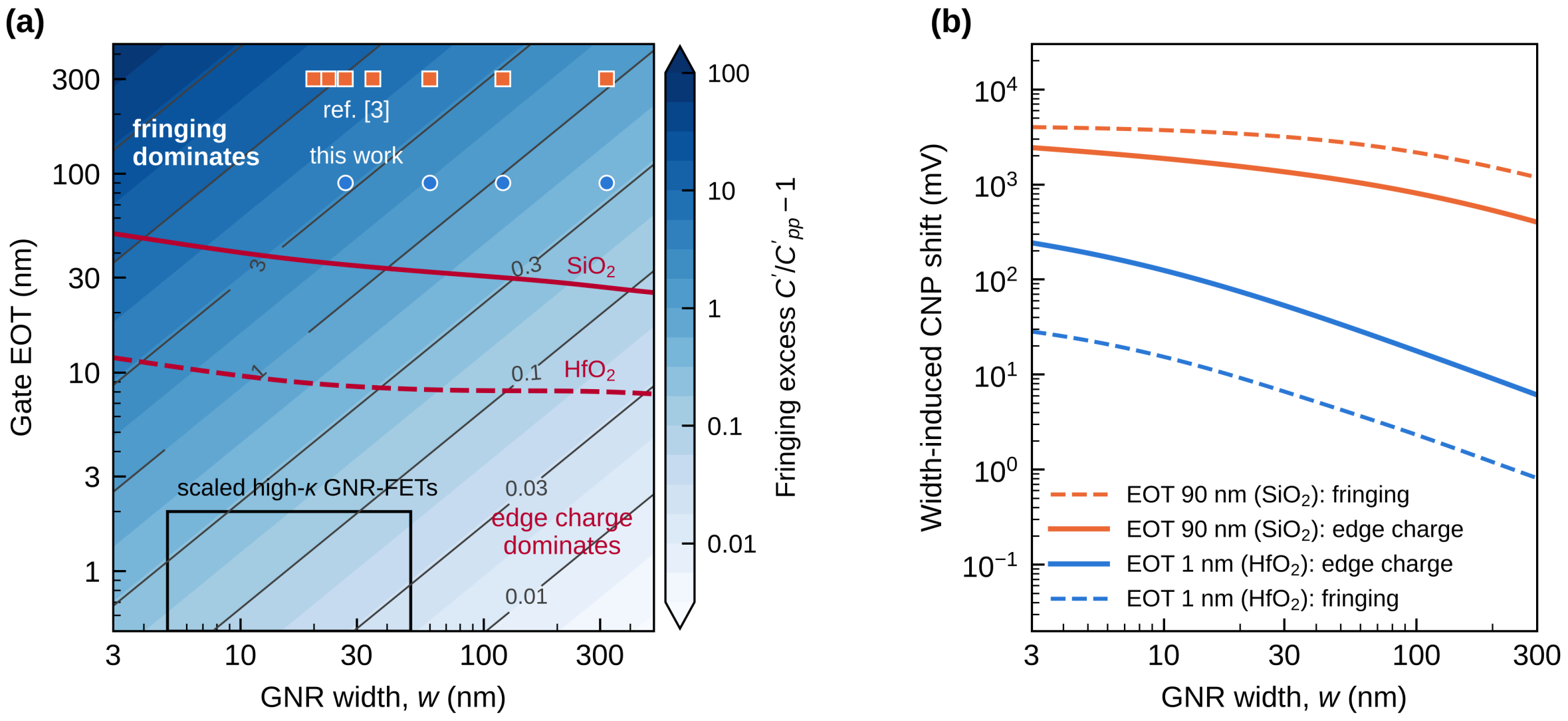

*Figure 13: (a) Fringing excess of the gate capacitance, $C'/C'_{pp} - 1$ (contours), for a ribbon of width w on a gate dielectric of equivalent oxide thickness EOT ($SiO_2$, air above). Red lines: R = 1, where the fringing and edge-charge contributions to the width-induced CNP shift are equal, for $|2\lambda/p_b|$ = 40 nm, on $SiO_2$ (solid) and on $HfO_2$ ($\varepsilon_r$ = 20, physical thickness 5.1 × EOT; dashed). Symbols: devices of this work and of reference [3]; box: scaled high-κ GNR transistors. (b) Width-induced CNP shift due to an edge charge of 0.2 e nm$^{-1}$ per edge (solid) and to fringing with $p_b = 10^{12}$ cm$^{-2}$ (dashed), for EOT = 1 nm ($HfO_2$) and EOT = 90 nm ($SiO_2$).*

Scaled GNR transistors, with high-κ gate dielectrics of EOT ≈ 1–2 nm [33] and widths of 5–50 nm, lie deep in the opposite regime (Figure 13(b)). For a $HfO_2$ gate with EOT = 1 nm (physical thickness ≈5 nm), fringing raises the capacitance of a 10 nm ribbon by ≈50% yet shifts its CNP by only ≈15 mV for $p_b = 10^{12}$ cm$^{-2}$, whereas an edge charge of 0.2 $e$ nm$^{-1}$, close to that found here, shifts it by ≈120 mV, falling to ≈60 mV at 27 nm and ≈18 mV at 100 nm. Two practical consequences follow. First, in thin-gate devices the width dependence of the threshold voltage is a direct and nearly model-independent measure of the edge charge, so the edge chemistry left by different etching and passivation processes can be compared quantitatively; thick back gates, by contrast, require the full electrostatic analysis above. Second, because $\Delta V_e$ varies almost as $1/w$, edge charge converts width variations into threshold-voltage variations: for the edge charge found here, $|dV_T/dw| \approx 10$ mV nm$^{-1}$ at $w$ = 10 nm, so 1 nm of line-width roughness produces a threshold-voltage spread of order 10 mV, which adds to the variability budget of ribbon transistors patterned top-down. At EOT ≈ 1 nm the quantum capacitance of graphene is comparable to the geometric capacitance; this broadens the transfer characteristic but, to first order, does not move the neutrality condition on which these estimates rely.

# 6 Conclusion

Combining device-scale electrostatic modelling, a quantitative analysis of the CNP on two oxide thicknesses and scanning gate microscopy, we have tested the proposal that the width dependence of the charge neutrality point in oxygen-plasma-etched graphene nanoribbons arises from n-type edges surrounding a p-type bulk. We have shown three results that argue against this hypothesis. First, fringing of the gate field alone, with uniform doping, reproduces the trend of the CNP shifting towards zero as the ribbon narrows, and produces a bias-dependent inversion of the carrier type at the edges that can mimic n-type edges in SGM measurements— a cautionary note for scanned-probe studies of any gated two-dimensional nanostructure. Second, once the electrostatic contribution is calculated exactly, the remaining edge charge of the present ribbons on 90 nm $SiO_2$ is acceptor-like (+0.24 $e$ $nm^{-1}$ per edge), as expected for oxygenated edges, whereas the crossover in the earlier devices on 300 nm $SiO_2$ requires a small donor-like contribution (−0.07 $e$ $nm^{-1}$); as no single edge-charge, field-coupled or contact-doping mechanism describes both, n-type edges are not an intrinsic consequence of oxygen-plasma etching. Third, SGM finds no n-type signature at the ribbon edges, excluding n-type edge regions wider than ≈10–20 nm. Because the two sets of devices also differ in bulk doping, passivation and fabrication batch, establishing whether that contribution depends on oxide thickness requires ribbons from a single fabrication run on two oxides. Fringing dominates the width dependence of the CNP whenever the gate dielectric is thicker than ≈30–50 nm; in scaled high-κ transistors the edge charge dominates instead, and the value found here would shift the threshold voltage of a 10 nm ribbon by ≈0.1–0.15 V. Fitting the CNP versus width with the exact strip capacitance, ideally on two (or more) gate-dielectric thicknesses, and SGM are simple tests that separate electrostatic from chemical edge effects in nanoribbons of graphene and other two-dimensional materials.